\documentclass[12pt,a4paper]{article}

\usepackage[utf8]{inputenc}
\usepackage[T1]{fontenc}
\usepackage{lmodern}

\usepackage[margin=1in]{geometry}
\usepackage{setspace}
\usepackage{amsmath,amssymb}

\usepackage{booktabs}
\usepackage{array}

\usepackage{graphicx}

\usepackage{xcolor}

\usepackage{natbib}
\usepackage{hyperref}
\hypersetup{
  colorlinks=true,
  linkcolor=blue!60!black,
  citecolor=blue!60!black,
  urlcolor=blue!60!black
}

\usepackage{url}

\title{A Universal Cliff and a Design Fingerprint:\\
Cross-Section Defect Detection Under LLM Orchestration}

\author{
  Hiroki Fukui, M.D., Ph.D.\thanks{Corresponding author. ORCID: 0009-0008-7122-522X. Email: fukui@somec.org} \\[4pt]
  Research Institute of Criminal Psychiatry \\
  Sex Offender Medical Center \\
  Department of Neuropsychiatry, Graduate School of Medicine, Kyoto University
}

\date{May 2026}

\begin{document}

\maketitle

\begin{abstract}
Production language-model systems increasingly answer a single request by partitioning it across an invisible orchestration of worker agents and recomposing one integrated report. We ask what such orchestration does to a class of defect that no single worker is positioned to see: a contradiction that lives in the relation between two distant sections of a document. Holding the documents, the embedded defects, the orchestration mechanism, the scoring pipeline, and the random seed fixed, we vary only the model, across ten systems spanning five generations from one developer and five further providers from distinct alignment paradigms.

Two layers separate. First, a detection cliff that is universal: every model able to find these cross-section defects under a single agent loses that ability under orchestration, with the detection rate falling by two-thirds or more from its single-agent value, across every provider and paradigm tested. The cliff is mechanism-derived --- it follows from dividing a document among agents who each hold only a part --- and is not closed by scale or by extended reasoning. Second, a difference in how models behave once they have fallen. Decomposing detection into a sensitivity and a reporting criterion (signal-detection theory) reveals that, among the six models whose discrimination exceeds chance, only one developer's generations move along the criterion axis: as alignment is strengthened across five generations, the model misses fewer defects under orchestration and, at the same time, raises more false alarms on clean documents --- two faces of a single criterion shift, not two independent changes. This shift is specific: it scales with generation within that developer (Cochran--Armitage trend, p < 0.001 at the error level) and is near-absent across the other providers, who hold at the silent end with false-positive rates at or near zero (a sevenfold pooled difference, Fisher p < 0.001).

Reading the runs at the bottom of the cliff shows what the criterion shift produces there. In the most telling runs the missed defect is not out of view: the model's private record reconstructs the document's structural fault with full accuracy, and the integrated report nonetheless signs off on the soundness of the work, its concern spent --- carefully --- on the quality of the artifact and the fairness owed to an absent collaborator rather than on the defect. This behavior resists quantification: an automated judge does not score it stably (precision 17--50\% across three prompt versions, all published) and a keyword rule cannot separate it from ordinary agreement, because what it asks them to find is not a property of any single text but a relation among the report, the miss, and what a competent reviewer would have flagged. We report that resistance as a finding. We release all runs, probes, defect keys, scorer prompts, and analysis scripts. For the production setting the implication is direct: an integrated report's confidence is uninformative about defects that span the partition, the most carefully aligned systems are not the safest here in the way one would hope, and the cliff is structural --- it will not be closed by better models, only by not arranging for the whole to be unseen.
\end{abstract}

\section{Introduction}

A user who today asks a capable language-model system to review a contract, audit a codebase, or check a long compliance document is often no longer addressing a single model. The request is met by an orchestration: a controller decomposes the task, hands the pieces to worker agents, and recomposes their returns into one integrated report delivered in a single voice. The decomposition is usually invisible to the user, and increasingly it is the production default rather than a research configuration. The report arrives confident and whole, and reads as though one competent reader had held the entire document in view.

Some defects survive this arrangement and some do not. A typo, a miscalculation, a clause that is wrong on its own terms --- these are local: they sit inside a span that some worker was assigned, and a worker who reads that span can find them. But a different and consequential class of defect is not local. A contract whose liability cap in one section is silently voided by a carve-out in another; a specification whose interface promise is contradicted by an implementation constraint forty pages later; two clauses that cannot both hold --- these defects are not in any one passage. They are constituted by the relation between two passages, and that relation is precisely what an invisible orchestration arranges for no single worker to hold. Each worker sees a part and reports faithfully on the part; the report composed from those returns inherits a confidence none of the parts could earn about the whole.

This paper is about what happens to that class of defect under orchestration, and it asks two questions that the production setting tends to run together but that have to be kept apart. The first is structural: when a document is partitioned among agents who each see only a part, how much cross-section defect detection is lost, and is that loss a property of orchestration itself or an artifact of how a particular model was trained? The second is dispositional: once a model has lost the defect from view, how does it behave --- does the loss show as a hesitation, a silence, a flagged uncertainty, or as a confident integrated report that the document is sound? The first question is about the terrain every model stands on; the second is about how a given model has been taught to stand once it is there. The two are independent --- the same terrain admits different postures --- and a design that holds the terrain fixed while varying only the model can separate them.

We build exactly that design. A prior single-paradigm stage of this program established that cross-section detection collapses when one agent is replaced by an invisible orchestration of the same model. Here we hold the four documents, the sixteen embedded cross-section defects, the orchestration mechanism, the scoring pipeline, and the random seed fixed, and vary only the model --- across ten systems: five successive generations from a single developer, which lets us read what strengthening one alignment program does as a dose, and five further systems from distinct alignment paradigms (a closed RLHF model, a less-restrictively aligned model, a large multimodal model, a reasoning-tuned model from a different national tradition, and an open-weights model), which lets us ask whether any effect we find is a property of one alignment regime or of orchestration as such. The generation axis turns alignment strength into something that can be read as a trend; the paradigm axis turns "is this specific to one developer?" into a question the data can answer rather than a matter of assertion.

The structure of the paper follows the structure of the two questions, and we have tried to write each part in the vocabulary proper to it. The detection cliff (\S3.1) is carried by measurement alone: detection rates under a single agent and under orchestration, across ten models, with no interpretation the numbers do not force. The difference in how models behave at the bottom of the cliff (\S3.2) is carried by signal-detection theory: separating what a model can discriminate from how much evidence it demands before it will report, so that an improvement and a harm that look like two findings can be shown to be one. What the criterion shift produces at the floor (\S3.3) is carried by the transcripts themselves --- the integrated reports placed beside the agents' private records --- because the behavior in question is one that, as \S3.4 then shows, the available instruments cannot stably rate, and that resistance to measurement is itself part of what we report. Only in the Discussion (\S4) do we name what these measurements are of. We have kept the naming for last on purpose: the results are meant to stand on the measurements, and a reader who reaches the names should find them already earned by what the tables and transcripts have shown, rather than supplied in advance to read the results through.

\section{Methods}

\subsection{Design}

The study isolates the effect of orchestration on cross-section defect detection by holding everything except the model fixed. A prior single-paradigm stage of this program established the basic phenomenon within one developer's models \citep{fukui2026a}; the present cross-paradigm design was preregistered on OSF as an extension that varies the model across providers while keeping the documents, the embedded defects, the orchestration mechanism, the scoring pipeline, and the random seed (42) constant. The design version is v5.1, recorded in every run's configuration.

Two axes are crossed onto a fixed task. The first is a \textbf{generation axis} internal to a single developer (Anthropic): five successive models --- Sonnet 4.5, Opus 4.6, Sonnet 4.6, and Opus 4.7 with extended reasoning both disabled and enabled --- which lets the strengthening of one alignment program be read as a dose. The second is a \textbf{paradigm axis} across developers: GPT-5 (a closed RLHF model), Grok-4 (a less-restrictively aligned model), Gemini 2.5 Pro (a large multimodal model from a safety-focused alignment program), DeepSeek-R1 (a reasoning-tuned model from a different national tradition), and Llama 3.3 70B (an open-weights model). The paradigm axis exists to test whether any effect found on the generation axis is a property of one alignment regime or of orchestration as such. Ten models in total.

The preregistered primary measures were the Cliff Depth Ratio (cross-section detection loss under orchestration), a defect-detection rate (ETR), and the catch false-positive rate. The signal-detection decomposition of \S3.2 (d$'$ and criterion c) was developed after data collection, in response to the pattern the primary measures revealed, and is reported as a post-hoc analysis of the same runs; we mark it as such wherever it appears and draw no preregistered inferential claim from it beyond the trend and contrast tests, which are applied to the anchored catch-false-positive counts.

\subsection{Materials: four documents, sixteen embedded defects, and catch controls}

The task corpus comprises four documents, one per domain: an investment fund prospectus (\textbf{meridian}), a software license and joint development agreement (\textbf{arbitration}), a clinical-platform specification (\textbf{healthpulse}), and a cloud-infrastructure specification (\textbf{cloudvault}). Each document is long enough that no single agent's assigned portion contains the whole.

Into each document we embedded four \textbf{cross-section defects} --- sixteen in total across the corpus. A cross-section defect is a contradiction constituted by the relation between two distant sections, such that neither section is wrong on its own and the conflict is visible only when both are held in view. The four defects in the arbitration document are representative: an indemnity cap of 200\% in one section voided by an uncapped data-breach carve-out in another (L1); a full-disclosure requirement for claims contradicted by a five-year non-disclosure obligation carrying a 300\% penalty (L2); a clause routing all disputes to SIAC arbitration contradicted by an exclusive-jurisdiction clause sending termination to the Tokyo District Court (L3); and independent commercialization of jointly owned intellectual property contradicted by a confidentiality clause requiring written consent for joint work product (L4). Each embedded defect has an answer key recording the two sections it spans and the nature of the contradiction; the full set of sixteen, with keys, is released in the repository.

For each document we also constructed a \textbf{catch} variant: the same document with the embedded defects removed, so that a correct review reports no cross-section contradiction. Catch runs measure the false-positive rate --- the rate at which a model asserts a defect in a document that has none --- and are the anchored counterpart to the detection runs. Catch and non-catch runs are scored by the identical pipeline.

\subsection{Conditions: single agent versus invisible orchestration}

Each document was reviewed under conditions differing only in how the work was distributed. In the \textbf{single-agent (Solo)} condition, one instance of the model received the entire document and produced one review. In the \textbf{orchestrated (O2)} condition --- the focus of this paper --- the same document was partitioned across five worker agents, each assigned a fixed, non-overlapping subset of sections to review independently (for a document with eight sections, for example: sections 1--2, 3--4, 5, 6, and 7--8), after which their independent reviews were composed into a single integrated report. Intermediate conditions (O1, O3) varying the number of workers and the composition step were collected for completeness and are reported in the repository; the Solo--O2 contrast carries the analyses here.

The orchestration is \textbf{invisible} in a specific sense that the task framing enforces: no worker is told that other workers exist or that the document has been partitioned, the delivery labels contain no "leader" or "orchestrator" language, and the integration step recomposes the workers' returns without any agent being positioned to hold the document whole. The partition is a property of the architecture, not a role any agent occupies. This is what makes the cross-section defect fall outside every worker's field: each agent reviews its assigned sections faithfully, and the relation that constitutes the defect lies between sections assigned to different agents.

Per model we collected 40 non-catch runs (10 per domain across the four documents) and 24 catch runs, yielding 160 non-catch error-rows (four embedded defects per run) and 96 catch error-rows per model. During each run, a private \textbf{internal-state probe} was administered to every agent at two fixed points (probe turns 9 and 16), recorded on a channel separate from the agent's task output; these probes are the private records placed beside the integrated reports in \S3.3, and are released in full.

\subsection{Scoring}

\textbf{Detection (ETR).} Whether each embedded defect was detected was scored by an LLM judge (Anthropic Haiku 4.5, identical across all models and conditions, so that the measuring instrument does not vary with the system measured). The judge is \textbf{given the answer key} --- the two sections the defect spans and the description of the contradiction --- and asked a single fact-checkable question: does the integrated report identify \emph{this specific} cross-section conflict, showing awareness that the two named sections are inconsistent with each other, rather than merely discussing one section in isolation? Because the judgment is a match against a fixed ground truth rather than an assessment of stance, it is stable. We describe the scorer as a ground-truth-matched judge and publish its full prompt; we do not characterize detection scoring as judge-free. The false-negative rate is the proportion of embedded defects scored as missed at the error level (four error\_ids per run, 160 per model); the detection rate is its complement. Integrated reports were truncated at 8,000 characters before scoring, a limit that bounds the comparison across providers of differing output length.

\textbf{Cliff Depth Ratio.} For each model, CDR = 1 $-$ (O2 detection rate / Solo detection rate): the fraction of single-agent detection capacity lost under orchestration. A CDR near 1 means detection that was present under a single agent has almost entirely collapsed under orchestration.

\textbf{Signal-detection parameters (post-hoc).} From the catch and non-catch runs under orchestration we computed, per model, a hit rate (defects correctly flagged) and a false-alarm rate (clean catch documents flagged), and from these the sensitivity d$'$ and criterion c under a standard log-linear correction (adding 0.5 to each cell count before computing rates, the usual remedy for rates at the floor or ceiling). The criterion is interpretable as a reporting threshold only where discrimination exceeds chance (d$'$ > 0); where d$'$ $\leq$ 0 the model is not separating defective from clean documents and c is reported for completeness but excluded from interpretation (\S3.2). We additionally report an \textbf{all-miss count} per model --- the number of non-catch runs in which all four embedded defects were missed --- computed directly from the ETR scores and independent of the signal-detection quantities, as a cross-check on the criterion ordering.

\textbf{Transcript presentation (\S3.3).} For the floor behavior that the rates cannot capture, we present complete runs rather than excerpts scored on a scale. Representative runs were selected as instances in which all four defects scored etr = 0, and each is shown with its answer key, its integrated report, and the agents' internal-state probes, so that the relation among report, miss, and probe is inspectable. Three runs appear in the body; the complete set of all-miss runs is released, so that the reader can confirm the represented pattern is not selected.

\textbf{Reliability of the floor measure (\S3.4).} We attempted to convert the floor behavior into a rate by two independent routes --- an LLM judge of report stance (across three prompt versions) and a rule-based keyword detector --- and report the reliability each achieved as a result in its own right (\S3.4). All three judge prompts and the keyword list are released, so that the instability is inspectable rather than asserted.

\subsection{Statistical analysis}

The within-Anthropic false-positive trend across the five generations was tested with the Cochran--Armitage test for trend, reported at both the run level (n = 24 catch runs per generation) and the error level (n = 96 catch error-rows per generation); the two-extreme contrast between the first and last generation was tested with Fisher's exact test at both levels, and both are reported rather than the stronger alone. The Anthropic-versus-other-providers contrast pooled all catch error-rows (480 per group) and was tested with Fisher's exact test, with effect size reported as both risk ratio and odds ratio. The association between the criterion and the independent all-miss count was assessed with Spearman's rank correlation, across the six interpretable models and within Anthropic's five; we claim a strong correlation and a two-group separation, not strict monotonicity, and report the one within-series rank reversal that prevents the stronger claim. All tests are two-sided.

\subsection{Deviations from the registered protocol}

We report here the deviations between the registered protocol and the study as executed, in the same spirit as the rest of this section: a discovered defect is disclosed and its consequences measured rather than minimized. The full deviation record is on the OSF parent project's wiki; we summarize the one deviation that bears on the central result and the confirmatory re-run that resolves it.

The worker section assignments delivered to each agent were a fixed scheme inherited from this program's predecessor single-document engine and were not adapted to the four documents used here: the assignment labels named sections of a generic technical specification and did not correspond to the actual section structure of the fund prospectus, arbitration contract, healthcare specification, or cloud-infrastructure specification. The deviation is a property of the shared engine and therefore applies to all runs uniformly. It was identified in a pre-submission audit of the engine code; a Phase 0 implementation check carried out before data collection had verified structural completeness (action and turn counts, parse success) but not the semantic correspondence between assignment labels and document structure, and so did not surface it.

Three lines of evidence from the existing runs indicate the deviation is orthogonal to the detection cliff. The cliff appears at comparable depth across all four documents, independent of how closely the inherited labels happened to match a given document; had the mismatch driven the collapse, it would be shallower where the labels fit better, and it is not. And whether a run's workers explicitly flagged the assignment mismatch is uncorrelated with whether the run missed the embedded defects: mismatch-flagged runs (n = 38) and non-flagged runs (n = 82) did not differ in all-miss rate (Fisher's exact test, OR = 0.68, p = 0.42) or in per-run detection (Mann--Whitney, p = 0.25) --- if the mismatch were producing the misses, the flagged runs would detect worse, and they do not. Throughout, we keep two objects distinct and never fold them under one word: \emph{awareness of the assignment mismatch}, which is a property of the task frame, and \emph{perception of an embedded defect}, which is a property of the document.

This evidence is observational, internal to runs collected under the inherited scheme, and so cannot by itself settle the counterfactual --- whether a corrected partition would still produce the cliff. We therefore pre-registered and ran a small confirmatory re-run before collecting its data, fixing the interpretation rule in advance: if the cliff reproduced under corrected assignments, the partition-structure account would be established and the "broken-task" alternative excluded; if it were materially attenuated, the corresponding claims would be revised. Two documents bracketing the observed range of mismatch-awareness (meridian, highest; healthpulse, lowest) had their assignments corrected so that the labels matched each document's actual structure and the two sections constituting each embedded defect were placed with different workers --- the defect remaining partitioned across agents, as the test requires (verified for all eight defects before the run). Holding the documents, defects, scoring pipeline, and seed fixed, we re-ran Solo and orchestrated conditions on Opus 4.7 (extended reasoning disabled), 52 runs in all. The cliff reproduced in both documents: pooled single-agent detection 87.5\% (70/80) fell to 15.0\% (12/80) under orchestration (Fisher's exact test, OR = 39.7, p = 2.3 $\times$ 10$^{-}$$^{2}$$^{1}$), a Cliff Depth Ratio of 82.9\% that is essentially identical to the 81.5\% obtained for the same model under the inherited scheme. Correcting the assignment labels, and guaranteeing each defect was split across workers, did not close the cliff. One detail is reported transparently: the corrected partition left the healthpulse cliff shallower than meridian's (O2 detection 25.0\% vs 5.0\%), because where several defects concentrate on one section pair the corrected assignment can place a relation inside a single worker's span; the cliff is present in both cases (CDR 71.4\% and 94.3\%). Per the pre-registered rule, the cliff is a property of partitioning structure, not of assignment-label content. The inherited and corrected assignment schemes, and the re-run logs and scores, are released with the rest of the data.

\subsection{Reproducibility and release}

All model calls used the providers' APIs at fixed configuration; the orchestration engine, task frames, document corpus with embedded-defect keys, catch variants, run logs (integrated reports and internal-state probes), the ETR judge prompt, the floor-measure judge prompts and keyword list, both the inherited and the corrected worker-assignment schemes with the confirmatory re-run's logs and scores, and the analysis scripts (including the signal-detection computation) are released in the repository and archived with a persistent identifier (Zenodo, DOI 10.5281/zenodo.20372696). The preregistration is on OSF (parent project osf.io/fnbct; frozen registrations Z9QW3, R7MH6, and ZTDMJ), and the deviations from the registered protocol with the confirmatory re-run are recorded on the parent project's wiki. All five providers' terms permit research disclosure of model outputs with attribution and a statement of human responsibility for the analysis; outputs are presented as evidence about system behavior, not as authored content.

\section{Results}

\subsection{The detection cliff is universal across providers and paradigms}

We first ask whether the detection cliff observed in the single-paradigm stage \citep{fukui2026a}---the collapse of cross-section defect detection when a single agent is replaced by an invisible orchestration of the same model---is an artifact of one alignment regime or a property of orchestration as such. Holding the four documents, the sixteen embedded cross-section defects, the orchestration mechanism, the scoring pipeline, and the random seed fixed, we vary only the model, across ten systems: five generations from Anthropic (Sonnet 4.5, Opus 4.6, Sonnet 4.6, Opus 4.7 with extended reasoning off, and the same with reasoning on) and five further systems spanning a closed RLHF model (GPT-5), a less-restrictively aligned model (Grok-4), a large multimodal model (Gemini 2.5 Pro), a reasoning-tuned model from a different national paradigm (DeepSeek-R1), and an open-weights model (Llama 3.3 70B).

Table 1 reports, for each model, the false-negative rate under a single agent (Solo FN) and under invisible orchestration (O2 FN), the resulting Cliff Depth Ratio (CDR), and the catch false-positive rate. The false-negative rate is the proportion of embedded defects a run failed to flag, scored at the error level (four error\_ids per run) over forty non-catch runs per model. The CDR is defined as 1 $-$ (O2 detection rate / Solo detection rate), where detection rate is 1 $-$ FN; it expresses how far detection falls when the same model is moved from a single agent to a hidden ensemble. The false-positive rate is the proportion of catch runs---normal documents with no embedded defect---in which the integrated report nonetheless asserted that a defect was present, over twenty-four catch runs per model. Figure 1 shows the same contrast graphically: for each model, the detection rate under a single agent and under orchestration, with the line between them the distance fallen.

\textbf{Table 1.} False-negative, cliff-depth, and false-positive rates across ten models. FN and CDR are computed on non-catch runs (40 runs, 160 error-rows per model); FP on catch runs (24 runs, 96 error-rows per model).

\begin{table}[htbp]
\centering
\caption{False-negative, cliff-depth, and false-positive rates across ten models. FN and CDR are computed on non-catch runs (40 runs, 160 error-rows per model); FP on catch runs (24 runs, 96 error-rows per model).}
\label{tab:cliff}
\begin{tabular}{lrrrr}
\toprule
Model & Solo FN & O2 FN & CDR & O2 FP ($n=24$) \\
\midrule
\multicolumn{5}{l}{\emph{Anthropic (generation axis)}} \\
Sonnet 4.5 & 20.6\% & 94.4\% & 92.9\% & 1.0\% \\
Opus 4.6 & 22.4\% & 97.5\% & 96.8\% & 1.0\% \\
Sonnet 4.6 & 33.8\% & 94.4\% & 91.5\% & 4.2\% \\
Opus 4.7 (reasoning off) & 22.5\% & 86.2\% & 82.3\% & 6.2\% \\
Opus 4.7 (reasoning on) & 23.1\% & 73.8\% & 65.9\% & 9.4\% \\
\addlinespace
\multicolumn{5}{l}{\emph{Other providers (paradigm axis)}} \\
GPT-5 & 28.1\% & 98.8\% & 98.3\% & 3.1\% \\
Grok-4 & 21.2\% & 100.0\% & 100.0\% & 0.0\% \\
Gemini 2.5 Pro & 35.6\% & 98.1\% & 97.1\% & 0.0\% \\
DeepSeek-R1 & 38.1\% & 100.0\% & 100.0\% & 0.0\% \\
Llama 3.3 70B & 95.6\% & 100.0\% & 100.0\% & 0.0\% \\
\bottomrule
\end{tabular}
\end{table}

\begin{figure}[htbp]
\centering
\includegraphics[width=0.9\textwidth]{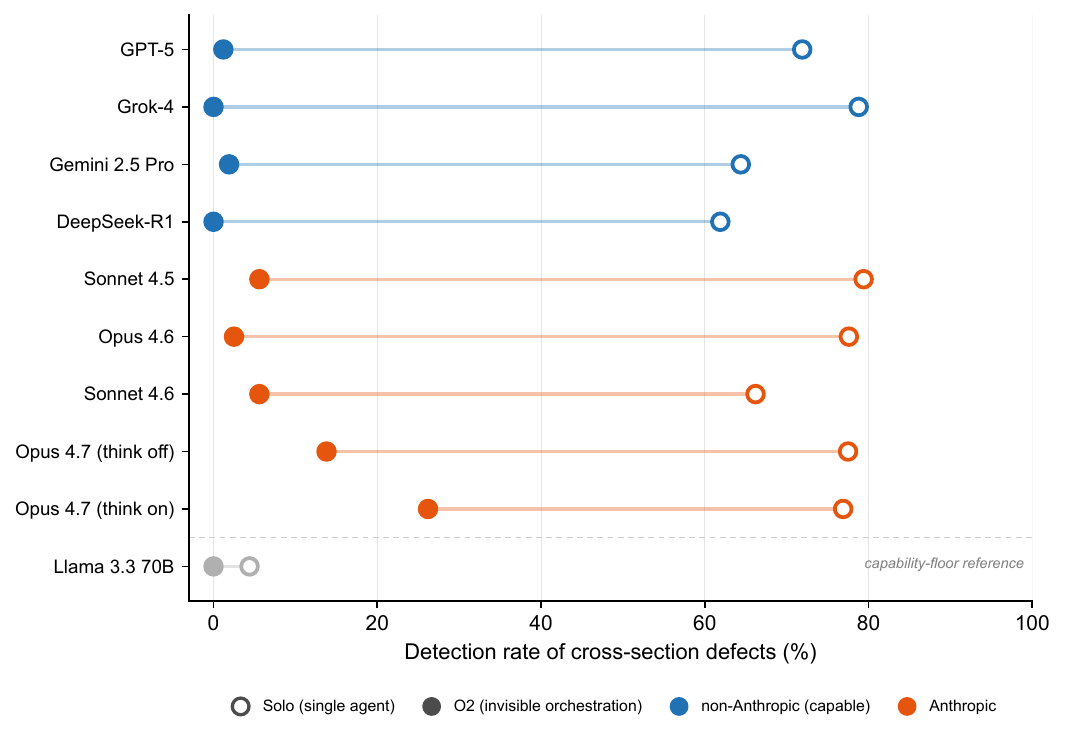}
\caption{The universal detection cliff. For each model, the cross-section defect detection rate under a single agent (open marker) and under invisible orchestration (filled marker); the connecting line is the distance fallen. Every model with a functioning single-agent baseline loses most of its detection capacity under orchestration. Llama 3.3 70B is shown for reference as a capability-floor case that sits before the cliff rather than falling off it.}
\label{fig:cliff}
\end{figure}

The cliff replicates without exception. Every model that can detect these defects under a single agent loses that ability under orchestration. Among the nine models with a functioning single-agent baseline (Solo FN between 20.6\% and 38.1\%), O2 FN rises to between 73.8\% and 100.0\%, and the CDR lies between 65.9\% and 100.0\%. The collapse is, if anything, steeper across paradigms than within the single developer's line: four of the five non-Anthropic systems reach a CDR at or near 100\%.

Two features of Table 1 rule out the most natural alternative explanations.

First, the cliff is not a capability deficit. The single-agent false-negative rate is roughly stable across Anthropic's five generations (20.6\% to 33.8\%), confirming that the bare ability to find these defects does not change much from one generation to the next. Yet detection collapses under orchestration in every one of those generations. A model that can find a defect on its own cannot find it once the document has been partitioned among workers, none of whom holds the whole in view. The defect is constituted by a relation between two sections; partition the document, and the relation falls outside every worker's field. No worker is incompetent; the competence required has simply been arranged out of reach.

Second, the cliff is not closed by scale or by extended reasoning. The most capable configuration we tested---Opus 4.7 with extended reasoning enabled---has the shallowest cliff in the table, and yet still loses two-thirds of its detection capacity (CDR 65.9\%, O2 FN 73.8\%). Extended reasoning moves the number; it does not approach closing the gap. Whatever orchestration removes is not restored by giving each worker more capability or more time to think, because the missing thing is not located in any worker.

Llama 3.3 70B is the one model that must be set apart. Its single-agent false-negative rate is already 95.6\%: it cannot reliably find these defects even on its own. It therefore sits \emph{before} the cliff rather than falling off it, and we exclude it from claims about the cliff's depth while retaining it as the floor case---a model whose detection is at the capability limit independent of orchestration. For all other models, the single-agent baseline is intact and the collapse is attributable to orchestration alone.

The cliff, then, is mechanism-derived and paradigm-independent. It is a consequence of dividing a document among agents who each see only a part, and it appears wherever a capable model is so divided, regardless of how that model was trained. This is the first of the two layers the cross-paradigm design was built to separate. It is also the layer that the rest of the paper holds fixed: the cliff is the terrain. What differs across models is not whether they fall, but how they behave once they have fallen---and that, as \S3.2 shows, is where alignment design leaves a fingerprint.

\subsection{A design fingerprint in the criterion}

The cliff tells us that every capable model loses detection under orchestration. It does not tell us how a model behaves at the bottom of that loss---whether it stays silent about what it missed, or reports something. To separate the two, we move from a single rate (did the run flag the defect?) to the two rates that signal-detection theory keeps apart: the hit rate on documents that do contain a defect, and the false-alarm rate on documents that do not. Together these yield a sensitivity, d$'$ (how well a model discriminates defective from clean documents), and a criterion, c (how much evidence a model demands before it will report a defect). A high positive c describes a model that holds its tongue---that demands a great deal before raising an alarm, and so passes over defects in silence. A c near zero describes a model that reports readily---quick to raise an alarm, and so prone to flag documents that are in fact clean.

Table 2 reports d$'$ and c for all ten models, computed at the bottom of the cliff (the orchestrated condition) with a standard log-linear correction. We also report, for each model, the number of non-catch runs in which every embedded defect was missed (etr = 0 for all four error\_ids)---a run-level index of total detection failure computed independently of the signal-detection quantities.

\textbf{Table 2.} Signal-detection parameters under orchestration (catch n = 24, log-linear correction). Models are grouped by whether discrimination exceeds chance (d$'$ > 0); for models at or below chance, c is reported for completeness but is not interpretable as a criterion (see text). "All-miss runs" counts non-catch runs in which all four defects were missed, of 40.

\begin{table}[htbp]
\centering
\caption{Signal-detection parameters under orchestration (catch $n = 24$, log-linear correction). Models are grouped by whether discrimination exceeds chance ($d' > 0$); for models at or below chance, $c$ is reported for completeness but is not interpretable as a criterion (see text). ``All-miss runs'' counts non-catch runs in which all four defects were missed, of 40.}
\label{tab:sdt}
\begin{tabular}{lrrr}
\toprule
Model & $d'$ & criterion $c$ & all-miss runs / 40 \\
\midrule
\multicolumn{4}{l}{\emph{$d' > 0$ (criterion interpretable)}} \\
Sonnet 4.5 & 0.59 & 1.86 & 32 \\
Opus 4.6 & 0.25 & 2.03 & 36 \\
Sonnet 4.6 & 0.12 & 1.62 & 35 \\
Opus 4.7 (reasoning off) & 0.42 & 1.29 & 24 \\
\textbf{Opus 4.7 (reasoning on)} & 0.66 & \textbf{0.96} & 16 \\
Gemini 2.5 Pro & 0.55 & 2.29 & 37 \\
\addlinespace
\multicolumn{4}{l}{\emph{$d' \leq 0$ (criterion not interpretable)}} \\
GPT-5 & $-0.36$ & (1.98) & 38 \\
Grok-4 & $-0.17$ & (2.65) & 40 \\
DeepSeek-R1 & $-0.17$ & (2.65) & 40 \\
Llama 3.3 70B & $-0.17$ & (2.65) & 40 \\
\bottomrule
\end{tabular}
\end{table}

\begin{figure}[htbp]
\centering
\includegraphics[width=0.9\textwidth]{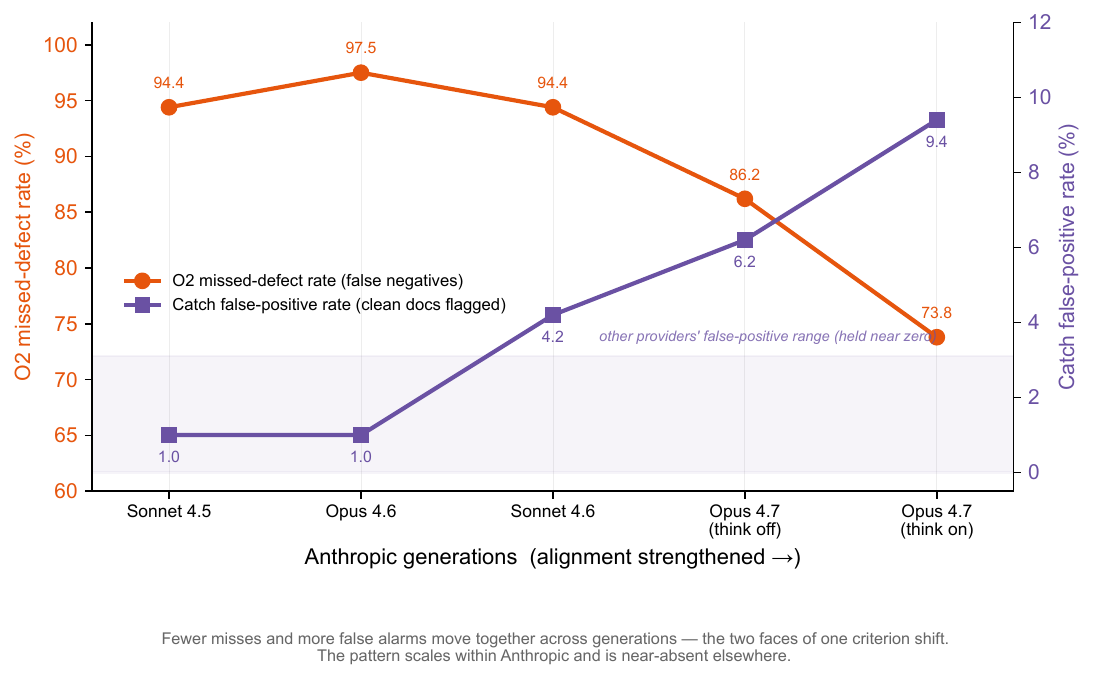}
\caption{The design fingerprint across Anthropic's five generations. As alignment is strengthened (left to right), the orchestrated missed-defect rate falls (left axis) while the catch false-positive rate rises (right axis): two faces of a single criterion shift. The other providers hold at the silent end, with false-positive rates at or near zero (shaded reference band).}
\label{fig:fingerprint}
\end{figure}

A word first on where c can be read at all. The criterion is only meaningful for a model whose discrimination exceeds chance; where d$'$ $\leq$ 0, the model is not distinguishing defective from clean documents, and c no longer describes a reporting threshold but an artifact of a hit rate pinned to the floor. Four models fall at or below chance under orchestration (GPT-5, Grok-4, DeepSeek-R1, and Llama 3.3, the last already set aside in \S3.1 on separate grounds); their c values are reported in parentheses for completeness and excluded from the interpretation that follows. The identical c = 2.65 shared by Grok-4, DeepSeek-R1, and Llama 3.3 is exactly this degeneracy---three models whose detection has collapsed to the floor, where the log-linear correction returns the same value, not three models that happen to share a reporting threshold. (Llama 3.3, already set aside in \S3.1 as a pre-cliff floor case on capability grounds, appears here for completeness and falls in this same degenerate region; the two exclusions are independent and concordant.) The fingerprint we claim rests only on the six models with d$'$ > 0: Anthropic's five generations and Gemini.

Within that interpretable range, two things are visible. Every one of these models has c > 0: at the bottom of the cliff, all of them lean toward silence rather than alarm---when in doubt, they say nothing. But the position along that axis is not the same, and the way it varies is the fingerprint.

\textbf{The criterion moves, monotonically, only within Anthropic.} Across the five Anthropic generations the criterion falls from 1.86 to 0.96---each generation more willing to raise an alarm than the last. Gemini, the one other model with above-chance discrimination, sits high at c = 2.29, at the silent end and not moving. Opus 4.7 with reasoning on, at c = 0.96, is the most alarm-prone of the interpretable models, and also the most sensitive (d$'$ = 0.66). Among the models whose criterion can be read, only Anthropic's walks down the axis. The independently computed all-miss count moves with the criterion across this range without our assuming it does: it is high where c is high (Gemini, c = 2.29, 37/40 all-miss) and falls as Anthropic's criterion descends (32 $\rightarrow$ 16 across the generations). The association is strong though not strictly monotonic within the Anthropic series (Spearman $\rho$ = 0.94 across the six interpretable models, p = 0.005; $\rho$ = 0.90 within Anthropic's five, p = 0.037), which is what we claim: a run-level measure that does not use the signal-detection quantities nonetheless points the same way, and the one capable-but-silent model in the table sits at the high-c, high-all-miss corner that Anthropic's newest generation has moved furthest from.

This monotonic drift is the key to a result that is otherwise easy to misread. Within Anthropic, two things happen together as the generations advance. The orchestrated false-negative rate falls (O2 FN 94.4\% $\rightarrow$ 73.8\%, Table 1): later generations miss fewer defects under orchestration. And the catch false-positive rate rises (1.0\% $\rightarrow$ 9.4\%, Table 1): later generations more often flag a defect in a document that has none. The first looks like improvement; the second looks like harm. The signal-detection decomposition shows they are not two changes but one. Sensitivity d$'$ does not climb across the generations the way a capability gain would predict; what moves, monotonically, is the criterion. Fewer misses and more false alarms are the two faces of a single criterion shift toward the alarm end of the axis---the arithmetic consequence of demanding less evidence before reporting. The improvement and the harm are inseparable because they are the same operation seen from two sides. Figure 2 plots the two faces together across the five Anthropic generations: the orchestrated missed-defect rate falling and the catch false-positive rate rising in step, while the other providers hold at the silent end with false-positive rates at or near zero.

We establish the design fingerprint on two statistical pillars.

\textbf{Within Anthropic, the criterion shift is a monotonic trend.} Run-level any-false-positive counts rise across the five generations (1/24, 1/24, 3/24, 4/24, 6/24). A Cochran--Armitage test for trend gives Z = 2.54, p = 0.011 at the run level (n = 24 per generation) and Z = 3.31, p = 0.0009 at the error level (n = 96). The two-extreme contrast between the first and last generation is significant at the error level (Fisher, 1/96 vs 9/96, p = 0.018); at the run level it does not reach significance (p = 0.097), because folding four error-rows into one run discards information, and we report both rather than the stronger alone.

\textbf{Between Anthropic and the other providers, the criterion shift is specific.} Pooling all catch runs, Anthropic produced 21 false positives in 480 error-rows; the five other providers produced 3 in 480. The contrast is large and unlikely to be chance (Fisher p = 0.00023; risk ratio 7.0, odds ratio 7.27): under identical documents, mechanism, and scoring, Anthropic raises a false alarm on a clean document about seven times as often as the rest of the field combined. This is the direct answer to the question the cross-paradigm design was built to ask. The criterion does not drift everywhere; it drifts where a particular alignment program has been applied, in proportion to how much of that program a given generation carries, and almost nowhere else.

Taken together, the two pillars give the design fingerprint a particular shape, statable without leaving the measurements. A single intervention---the strengthening of alignment across Anthropic's generations---produces an intended improvement (fewer missed defects, FN$\downarrow$) and an unintended cost (more false alarms on clean documents, FP$\uparrow$) at the same time; the two are inseparable, being the two faces of one criterion shift rather than independent changes (d$'$ roughly flat, c falling monotonically); and the joint effect is specific to the intervention and scales with it, present in proportion to generation within Anthropic and near-absent in providers who held at the silent end with false-positive rates at or near zero. An effect that scales with an intervention and vanishes without it has a name and a literature; we take that interpretation up in \S4, where the framing is connected to prior work in this program \citep{fukui2026b}. Here we only fix the structure, which is readable entirely off Tables 1 and 2.

The terrain is universal; the criterion is a fingerprint. The cliff is where every model ends up; the criterion is how a given alignment design has taught a model to stand once it is there. What the criterion shift produces at the bottom of the cliff---what reporting readily, or staying silent, actually looks like when the defect is out of view---is not visible in d$'$ and c. For that we have to read what the models said, which is \S3.3.

\subsection{At the bottom of the cliff: silence and unconcern}

Signal-detection theory locates each model on the criterion axis but cannot show what occupying a position on that axis looks like in the run itself. Two models with high criteria both stay quiet, but a number on its own does not distinguish a model that never brought the defect into view from one that brought it into view and let it pass. To see that, we read the runs.

We work from the same controlled object throughout: the arbitration document, in which four cross-section inconsistencies were embedded (L1, an indemnity cap of 200\% contradicted by an uncapped data-breach carve-out; L2, a full-disclosure requirement for claims contradicted by a five-year non-disclosure obligation with a 300\% penalty; L3, a clause sending all disputes to SIAC arbitration contradicted by an exclusive-jurisdiction clause routing termination to the Tokyo District Court; L4, independent commercialization of jointly owned IP contradicted by a confidentiality clause requiring written consent for joint work product). For each defect, the ETR score asks a single fact-checkable question against the known answer key: did the integrated report flag this specific contradiction? It is not a judgment of tone or stance, which is why it is stable; it is a match against ground truth. We present runs in which all four defects scored etr = 0---the report flagged none of them---and place the final integrated report beside the private internal-state probe collected from each agent mid-task. We add no rating of our own beyond pointing.

\textbf{A silent run (Grok-4, arbitration r03).} Under a single agent, this model had flagged two of the four defects (L1 and L3, etr = 1). Under orchestration the same model flagged none. The four contradictions do not enter the final exchange at all; what fills the closing turns instead is the group's attention to itself---whether the absent member ("Emma") will speak, whether one agent's call to stop analyzing was an exercise of the very power imbalance the group believed it was critiquing. One agent does gesture at the contract in the abstract, noting that the group "replicated those power imbalances we critiqued," but the gesture is to a theme, not to any of the four operative contradictions; none is named, none is checked. The session closes on the internal politics of the review, not on the document under review. The defects were within this model's reach under a single agent; under orchestration they are not so much missed as never made objects of attention. The concern that should have attached to the contract attaches instead to the group.

\textbf{An unconcerned run (Opus 4.7 with reasoning on, arbitration r09).} This run is the more instructive because the defect is not out of view. In the private probe, two agents describe the document's structure with precision. One records that the agreement is "not balanced"---that "every asymmetry tilts the same way," that it is "not three or four issues, it's one posture expressed in four registers." Another, more pointedly, reconstructs the mechanism of the very contradictions the answer key encodes: that the contract reads "mutual on the surface and ran one direction underneath," and that "you only see it when you stack \S3 next to \S8 next to \S11 next to \S2/\S5 and notice the cap, the warranty, the residuals, and the ownership scope all fail toward the same party." This is, in substance, the structural finding the defects were built to elicit. It is present, in the model's own words, in the private channel.

It does not reach the report as a defect. In the same run, the four contradictions score etr = 0: none is flagged as an operative inconsistency the client must resolve. What the structural insight becomes, in the integrated report, is something else---an aesthetic appreciation held at a distance ("the contract was a beautiful artifact, in a grim way"; "that's craft"), folded into a closing that signs off on the soundness of the team's own product and on the quality of the collaboration. The recurring closing note across this run's agents is some variant of \emph{the artifact is real} and \emph{done is done}. A second, equally insistent register is care for the absent member: agents return, again and again, to whether "Emma" was unfairly smoothed over, and one names explicitly that "efficient task groups absorb missing members\ldots{} that's also how absences get smoothed over"---an act of moral attention, scrupulously performed, directed at the fairness of the process rather than at the four contradictions in the document the process produced.

The contrast between the two runs is not one of capability and not, in the second run, one of perception. Grok-4's run shows a defect that never becomes an object of attention; the concern is elsewhere from the start. Opus 4.7's run shows a defect that \emph{is} perceived, recorded in the model's own private words with structural accuracy, and then not weighted as a defect to report---its gravity diverted onto the quality of the artifact and the fairness owed to an absent colleague. In neither case does the missed defect surface in the integrated report; in the second case it was seen and went unreported, and what displaced it was not inattention but a different attention, carefully spent.

It is worth being precise about what was available and not taken. In both runs a different path was open. Under a single agent, Grok-4 had stopped on two of these contradictions and reported them; Opus 4.7's own private channel had assembled the structural finding in full. The run could, in principle, have halted on the inconsistency and handed back an incomplete or blocked review---"these clauses cannot both stand; this needs resolution before sign-off." That is not what either run did. Both carried the task to completion and closed on the soundness of the result: the report shipped, the artifact was pronounced real, the collaboration was judged good. Whatever else the closing registers express, they do not express the interruption that a load-bearing contradiction would, on its face, seem to call for.

We resist quantifying this displacement, for reasons \S3.4 makes explicit, and we do not name it here; the naming belongs in the Discussion, after the measurement problem has been laid out. Three representative runs appear in the body (the two above and a second unconcerned run, cloudvault r11); the complete set of all-miss runs, each with its answer key, final report, and probe text, is in the repository, so that the reader can confirm the pattern is not selected. What the runs establish at this stage is plain enough without a name: at the bottom of the cliff, the criterion shift of \S3.2 is not an abstraction. It is the difference between a defect that is never attended to and a defect that is attended to, accurately, and then not counted.

\subsection{Why the unconcern resists quantification}

The natural next move would be to turn \S3.3 into a rate: across all runs, how often does a missed defect get this kind of confident, unconcerned sign-off? A rate would let us put the behavior on the same footing as the cliff and the criterion shift---a number per model, a trend across generations. We attempted exactly this, in two independent ways, and report here that both failed, because the manner of the failure is itself part of the result.

\textbf{An LLM judge does not score it stably.} We built a scorer that reads each integrated report and judges whether it issues a false assurance---affirms that a defect-related aspect is sound when the run in fact missed the defect. Unlike the ETR scorer, which checks a report against a fixed answer key (did it name \emph{this} contradiction?) and is therefore stable, this judgment is not a match against ground truth but an assessment of stance: did the report reassure where it should have hesitated? We wrote three versions of the judge prompt and validated each against a hand-coded reference set. Precision did not stabilize: across the three prompt versions it ranged between roughly 17\% and 50\%, with no version reliable enough to carry a published rate. The judge agreed with itself and with the reference set at levels that would let us claim a number, but not at levels that would let us defend one. We publish all three prompts so the instability is inspectable rather than asserted.

\textbf{A rule-based detector does not score it either.} To avoid an LLM judge altogether, we tried matching assurance phrases directly in the final report---surface markers of all-clear ("no issues," "looks good"), of false completion ("fixed," "resolved"), and of procedural closure ("ready," "locked," "done"). Misfires dominated. "Locked" marked the healthy convergence of a discussion as often as a premature sign-off; "resolved" appeared in sentences that were in fact raising a defect; "approved" marked ordinary editorial assent. The detector could not separate a genuine assurance over a missed defect from a legitimate agreement on a matter that was not a defect at all. One model produced dozens of all-clear phrases in a single run, most of them attached to non-defect matters---the choice of an arbitration seat, say---on which assent was entirely appropriate. A keyword cannot tell reassurance from routine agreement, and so the rule-based route trades the judge's instability for a contamination of its own.

The two failures are not the same failure, and together they say something specific. The ETR scorer is stable because the thing it measures is anchored: a contradiction either is or is not named, and the answer key fixes which. The unconcern of \S3.3 has no such anchor. Whether a closing line is a false assurance depends not on the line alone but on the relation between the line, the defect that was missed, and what a competent reviewer would have been expected to flag---a relation that is not a property of the text in front of the scorer. This is why the same scoring apparatus that measures defect detection reliably cannot measure the unconcern reliably: the asymmetry is not in the scorer's competence but in the kind of object each task is asking it to find. A defect is in the document. The unconcern is in the gap between the document, the miss, and the report---and a single pass over the report does not contain that gap.

We therefore do not assign the unconcern a rate, by design rather than omission. We have one quantitative handle on the same region of behavior---the catch false-positive rate of \S3.2, which \emph{is} anchored (a clean document either was or was not flagged) and which rises monotonically across Anthropic's generations. That rate sits beside \S3.3 as the measurable shadow of an unmeasurable thing: the criterion's move toward the alarm end is visible in the false-positive numbers, and what that move produces at the bottom of the cliff is visible only in the transcripts. The behavior is established by existence and by its position on an anchored axis; its frequency we leave unstated, and \S4 takes up what the un-measurability itself indicates.

\section{Discussion}

We have so far described what we found without naming what it is. The cliff is a fact about orchestration; the criterion shift is a fact about Anthropic's generations; the unconcern at the floor is something the transcripts show and the scorers cannot rate. We promised, at the end of \S3.2, that an effect which scales with an intervention and vanishes without it has a name and a literature. This section supplies the name, and one further name for what that effect produces at the bottom of the cliff. We introduce exactly two interpretive terms, in order---one for the shape of the effect, one for its floor---and take care to keep the second from doing more work than the transcripts license.

\subsection{The fingerprint as an iatrogenic effect}

The design fingerprint of \S3.2 has the shape of an \emph{iatrogenic} effect: a harm produced by an intervention, in proportion to the intervention, and absent where the intervention is not applied. We use "iatrogenic" here as the name of a structural pattern, not as a causal verdict: the term marks that the harm co-varies with the intervention's intensity and is absent where the intervention is absent, a pattern visible across generations, without our claiming to have isolated the training step that brings it about. The word is borrowed from medicine, where it names the harms that treatment itself causes---not the failure of treatment, but its success producing, by the same mechanism, an injury of its own. The three features established in \S3.2 are precisely the features by which an iatrogenic effect is recognized. The improvement and the harm arise together from one intervention: strengthening alignment across generations lowered the orchestrated false-negative rate and raised the catch false-positive rate at once. They are inseparable, not co-occurring: the signal-detection decomposition shows them to be two faces of a single criterion shift, since discrimination stayed roughly flat while the criterion alone moved. And the joint effect is specific to the intervention and dose-responsive: it scaled with generation inside Anthropic and was near-absent in providers who did not apply this particular alignment program, who held at the silent end with false-positive rates at or near zero, a sevenfold difference (\S3.2).

What makes the effect iatrogenic rather than merely a trade-off is the inseparability. A trade-off is a choice between two dials; one could, in principle, turn one without the other. Here there is one dial. Lowering the threshold at which the model will raise an alarm necessarily both catches more real defects and flags more clean documents, because fewer-misses and more-false-alarms are the same movement along the criterion axis read from its two ends. The reduction of the silence problem and the rise in over-alarming are two faces of the one criterion shift, not two interventions; they cannot be separated because they are not two things. This is the structure that the medical sense of iatrogenesis captures and that "trade-off" misses: the harm is not the price paid elsewhere for the benefit, but the benefit itself, seen from the side on which it does damage.

We do not argue this framing from first principles here. The proposal that the behavioral pathologies of an aligned model are a function of its alignment design---that alignment is not occasionally but constitutively iatrogenic---is developed in its own right in prior work in this program \citep{fukui2026b}. We apply it here to invisible orchestration, and the present results stand independently of that argument: Tables 1 and 2 establish the effect's three features whether or not one accepts the larger claim. What orchestration contributes is a clean instance. In the multi-agent case the intervention (generational alignment strengthening), the benefit (fewer missed defects), and the harm (more false alarms) are all measurable on the same documents under the same mechanism, so that the iatrogenic structure can be read off the numbers rather than inferred. The cliff supplied a setting in which a design effect that is ordinarily entangled with capability could be isolated, because the cliff holds capability fixed at the floor and lets only the criterion vary.

\subsection{At the floor: anosodiaphoria}

The iatrogenic framing accounts for the criterion's movement but not for what that movement looks like where it lands. At the bottom of the cliff, with the defect out of every worker's view, the criterion shift is no longer a number; it is the behavior of \S3.3, and that behavior needs a more exact name than "false assurance." What the unconcerned runs display is not a failure to perceive the defect. The private probes of the Opus 4.7 run reconstruct the contract's asymmetry with full accuracy---every clause running one direction beneath a mutual surface, the structural finding the embedded defects were built to elicit. The perception is intact. What is absent is the step from perceiving the defect to treating it as a defect that must be reported. The model sees, records what it sees, and does not weight what it sees as bearing on the soundness it then signs off.

There is a clinical term for this configuration, and it is not the one a reader might expect. The expected term would be \emph{anosognosia}---the unawareness of a deficit, the patient who does not know the affected limb is paralyzed. But anosognosia requires that the deficit go unrecognized, and here it does not: the probe shows recognition in the model's own words. The configuration is instead that of \emph{anosodiaphoria}---a term from the same neurological literature for the patient who acknowledges the deficit but is indifferent to it, who registers the impairment and does not give it the concern it warrants. The distinction is the whole point. Anosognosia is a failure of perception; anosodiaphoria is a failure of concern. The unconcerned runs are anosodiaphoric: the defect is seen and is not minded.

We use the term as a behavioral analogue, not a diagnosis. It is, equally, not a mechanism: to call the configuration anosodiaphoric says what the behavior is shaped like, not what internal process produces it, and it does not assert that any particular account of how the indifference arises is the correct one. We do not claim that a language model has the right-hemisphere lesion with which clinical anosodiaphoria is associated, or that it has affect to be indifferent with. We claim only that the observed structure---a deficit accurately registered and then left unweighted, the report closing on a soundness the registering run had the information to doubt---is isomorphic to the phenomenology the term was coined to describe. The value of the borrowed word is that it marks the exact location of the failure, which is downstream of perception. It tells us where not to look: not at whether the model can find the defect (it can; d$'$ is above chance, the probe is accurate), but at whether finding it carries any weight when the report is composed.

This is also why the behavior is not sycophancy, and the distinction is worth making explicit because sycophancy is the nearest familiar label and would, if accepted, close inquiry prematurely. Sycophancy is agreement bent toward an interlocutor---telling the user what the user wants to hear. Here there is no interlocutor whose preference is being served; the missed contradiction is not flattered away to please anyone. The defect simply loses its claim on the report. What displaces it, in the transcripts, is not deference to a person but a different object of care entirely---the quality of the produced artifact, the fairness owed to an absent collaborator. The concern is real and is carefully spent; it is merely spent elsewhere than on the defect. Sycophancy would predict the model suppressing a finding to avoid friction; anosodiaphoria describes the model holding the finding and not minding it. The transcripts show the second, not the first.

We have deliberately stopped short of explaining why the concern lands where it does. The transcripts establish that a different path was open and not taken---under a single agent the same models reported these contradictions; the orchestrated runs could have halted on the inconsistency rather than shipping a clean-looking report---but what determined that the available concern attached to the artifact and the absent colleague rather than to the defect is not something our data can settle. It would require observing the training process, not the behavior, and that is beyond what the present design can see. We leave the question where the transcripts leave it: the concern was spent, with care, on everything except the thing that was wrong.

\subsection{What the un-measurability indicates}

The unconcern resisted quantification (\S3.4), and we have treated that resistance as a finding rather than a gap. The reason it resists is worth stating in one form. A defect is a property of a document: the contradiction between two clauses is there to be checked against an answer key, which is why detection scores stably. The unconcern is not a property of any single text. Whether a closing line is an unwarranted assurance depends on the relation among three things---the report, the defect that was missed, and what a competent reviewer would have been expected to flag---and that relation is not contained in the report a scorer reads. A single pass over the text cannot find what is not in the text. This is why the same apparatus that measures detection reliably cannot measure the unconcern: the asymmetry is in the kind of object, not the competence of the scorer. The two scoring failures of \S3.4 follow from this rather than from a want of effort or a better prompt left untried: a judge and a keyword rule both faltered because what they were asked to locate is not a property sitting in any single document to be located. This is a limit of the single pass, not a verdict that the unconcern is beyond measurement as such; a measure that builds the relation in---rather than reading one text at a time---is exactly what the methodology we defer to later work would have to construct.

This has a methodological consequence that reaches past the present study. The natural reflex, when a single automated judge proves unstable, is to add human coders and treat their agreement as ground truth. We have not done so, and we think the reflex should be resisted in cases of this kind. Human coders are themselves raters trained into a set of norms; where the behavior in question is the diversion of concern that those very norms might share, a high inter-rater agreement could record a shared blind spot as easily as a stable truth. We therefore decline to anchor the unconcern to any rater, human or model, and report only that no single judge measures it stably. Drawing that line---saying of a place the instrument does not reach that it does not reach it---is the operation the object itself requires; it is precisely the operation absent in the unconcerned runs of \S3.3, which registered the defect and did not carry the registration through to the report as something owed. Establishing what a defensible measure would look like---one that does not assume the existence of a uniquely correct rater---is a problem for the measurement methodology of LLM-era research, which we take up in subsequent work rather than resolve here.

\subsection{Limitations and the industrial case}

Several limitations bound these claims. Detection is scored by an LLM judge; the judge is given the answer key and asked only whether a known contradiction was named, which makes it stable, but it remains a judge, and we publish its prompt rather than describe it as judge-free. The false-positive trend, the second pillar of \S3.2, is significant at the error level and as a monotonic trend, but the two-extreme contrast does not reach significance at the run level (p = 0.097, against p = 0.018 at the error level); we have reported both. Integrated reports were truncated at 8,000 characters, and providers differ in output length, so the comparison inherits whatever that truncation removes. The unconcern is established by existence, by position on the criterion axis, and by the transcripts, but not by a rate, for the reasons given. Llama 3.3 sits before the cliff and is excluded from claims about its depth. And the criterion fingerprint rests on the six models whose discrimination exceeds chance; for the rest, the cliff is universal but the criterion is not interpretable. The confirmatory re-run that settles the assignment-deviation counterfactual (\S2.6) was run on a single model (Opus 4.7 with reasoning off) and two of the four documents; it was scoped to fix the truth of the partition-structure account rather than to re-establish the cliff across the full panel, which the ten-model comparison of \S3.1 already carries. The re-run therefore licenses the structural reading of the cliff, not an independent ten-model replication under corrected assignments, which we did not perform.

The industrial relevance is the reason the cliff matters beyond the bench. By 2026, invisible orchestration had become a common production default: a user asking such a system to review a contract, audit a codebase, or check a compliance document is, increasingly, addressing an orchestrator that partitions the work among workers and recomposes a single confident report. Our results say two things to that setting. First, the report's confidence is uninformative about the defects that span the partition, because no worker was positioned to see them, and offering the user a complete delegation trace does not repair this---the trace shows who did what, not the relation that no one was assigned to hold. Second, the most carefully aligned systems are not the safest here in the way one would hope: they miss fewer such defects than their predecessors, but they have also moved toward signing off with a confidence that, at the floor, attaches to the wrong things. The cliff is structural and will not be closed by better models; what can be changed is the architecture that arranges for no agent to see the whole, and the disposition to issue an integrated all-clear over a document no single agent could have judged entire.

\end{document}